\documentclass[pra,amsmath,amssymb,superscriptaddress,twocolumn]{revtex4-1}
\usepackage{dcolumn}
\usepackage{bm}
\usepackage{bbm}
\usepackage{amssymb, amsbsy, amsmath, latexsym, dsfont, array, layout, graphicx,mathrsfs,color}
\usepackage{makecell,multirow}
\usepackage[colorlinks,linkcolor=blue,anchorcolor=red,citecolor=red]{hyperref}

\newcommand{\one}{\mathds{1}}
\newcommand{\ket}[1]{\left|{#1}\right\rangle}
\newcommand{\bra}[1]{\left\langle{#1}\right|}

\newcommand{\ketbra}[2]{\left|{#1}\rangle\!\langle{#2}\right|}

\usepackage{array}
\newcommand{\PreserveBackslash}[1]{\let\temp=\\#1\let\\=\temp}
\newcolumntype{C}[1]{>{\PreserveBackslash\centering}p{#1}}
\newcolumntype{R}[1]{>{\PreserveBackslash\raggedleft}p{#1}}
\newcolumntype{L}[1]{>{\PreserveBackslash\raggedright}p{#1}}

\begin{document}

\title{Experimental entropic test of state-independent contextuality via single photons}
\author{Dengke Qu}
\affiliation{Beijing Computational Science Research Center, Beijing 100084, China}
\affiliation{Department of Physics, Southeast University, Nanjing
211189, China}
\author{Pawe{\l} Kurzy\'{n}ski}
\affiliation{Faculty of Physics, Adam Mickiewicz University, Umultowska 85, 61-614 Pozna\'{n}, Poland}
\author{Dagomir Kaszlikowski}
\affiliation{Centre for Quantum Technologies, National University of Singapore, 3 Science Drive 2, 117543 Singapore, Singapore}
\affiliation{Department of Physics, National University of Singapore, 3 Science Drive 2, 117543 Singapore, Singapore}
\author{Sadegh Raeisi}
\affiliation{Department of Physics, Sharif University of Technology, Tehran, Iran}
\author{Lei Xiao}
\affiliation{Beijing Computational Science Research Center, Beijing 100084, China}
\author{Kunkun Wang}
\affiliation{Beijing Computational Science Research Center, Beijing 100084, China}
\author{Xiang Zhan}
\affiliation{School of Science, Nanjing University of Science and Technology, Nanjing 210094, China}
\author{Peng Xue}\email{gnep.eux@gmail.com}
\affiliation{Beijing Computational Science Research Center, Beijing 100084, China}

\date{\today}

\begin{abstract}
Recently, an inequality satisfied by non-contextual hidden-variable models and violated by quantum mechanics for all states of a four-level system has been derived based on information-theoretic distance approach to non-classical correlations. In this work, we experimentally demonstrate violation of this inequality with single photons. Our experiment offers a method to study a distinction between quantum and classical correlations from an information-theoretic perspective.
\end{abstract}

\maketitle

Quantum theories~\cite{S60,B66,KS67,M93,C08,BBCP09} provide advantages over classical ones for certain communication~\cite{BB84} and computational tasks~\cite{Shor}. Classical and quantum information processing scenarios differ on a fundamental level~\cite{NC00}, yet the outcomes of both types of scenarios are always classical. If one performs a test $X$ with outcomes $\{x\}$, the information content of the test's statistics can be quantified via Shannon entropy $H(X)=-\sum_x P(X=x)\log_2P(X=x)$, no matter whether the tested system was classical or quantum. In order to detect non-classicality in such tests, one should understand what the typical classical properties, revealed by Shannon entropies, are and how quantum ones differ. It is known that quantum correlations are stronger than classical ones, but when it comes to entropies of quantum tests, the differences are much more sophisticated. For example, non-classical features of entropies coming from quantum experiments can be amplified if one post-processes measured data, e.g., combines non-classical and classical probability distributions~\cite{C13}.

Contextuality is one of the major differences between quantum and classical physics~\cite{KS67}: it states that measurement results of some physical property may depend on how this property is measured. Previous tests of contextuality focused mainly on the probability distribution of measurement results. Recently, the entropic tests of quantum contextuality were introduced~\cite{KRK12,CF12,CF13} and further investigated experimentally~\cite{ZKK+17}. However, these tests are state-dependent, i.e., departure from classical behavior can be detected only if the system is prepared in some special state. This paradigm was changed in~\cite{RKK15} where the entropic approach to the state-independent contextuality was proposed allowing non-classical features to be observed for any state of the system. This is done with the help of a newly discovered multi-partite information-theoretic distance for binary  measurements with two outcomes $\pm1$. This distance measure uses Shannon entropy and yields a state independent, multi-partite non-contextual inequality that resembles the correlation-based inequality~\cite{C08,ARBC09,KZG+09,ZWD+12}. In this paper we demonstrate its violations with single photons.

\begin{figure}
  \includegraphics[width=0.35\textwidth]{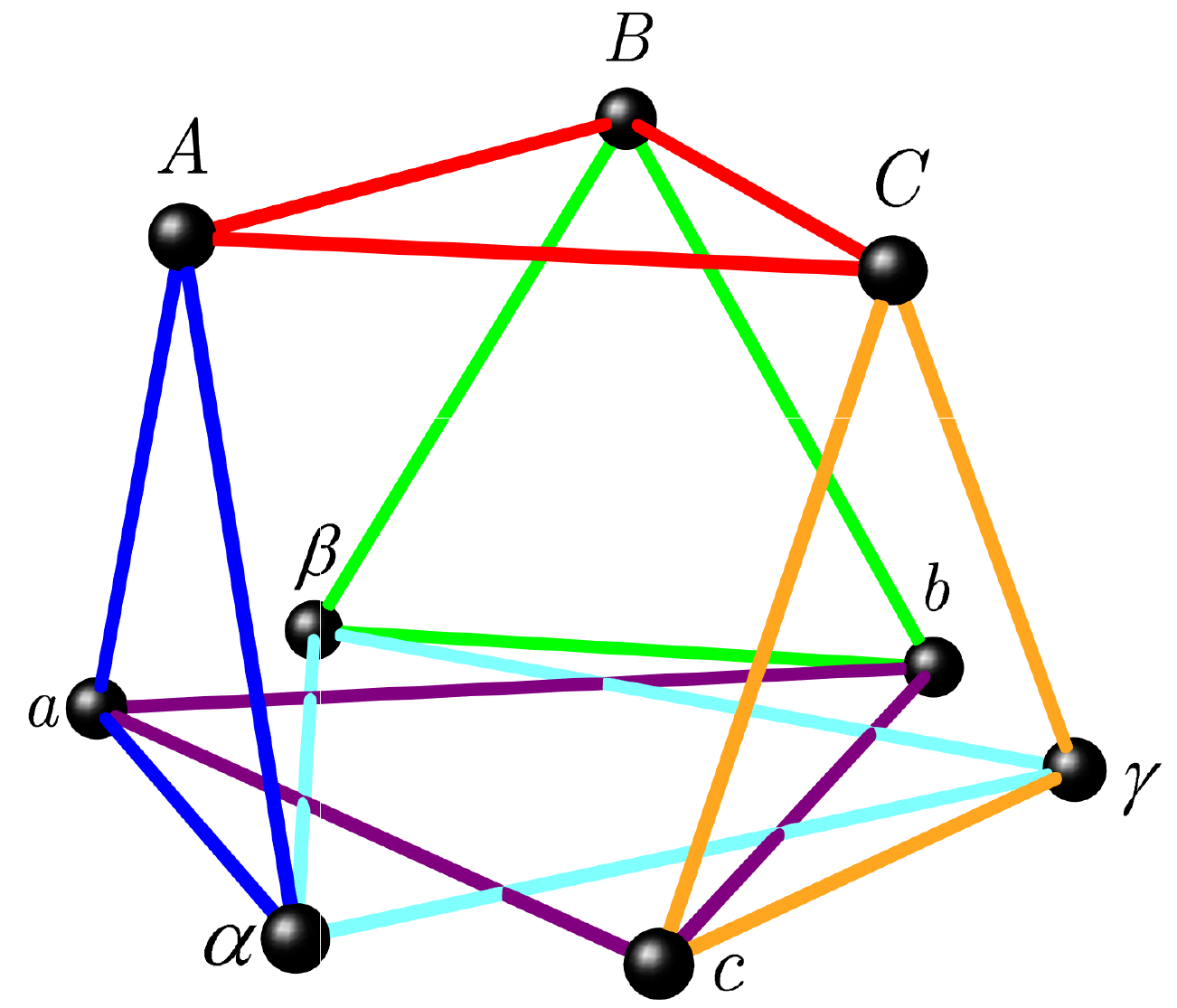}
  \caption{Illustration of compatibility relations among the nine binary-outcome observables. Compatible observables are connected by links of the same color.}
  \label{fig:1}
\end{figure}

\begin{figure}
  \includegraphics[width=0.4\textwidth]{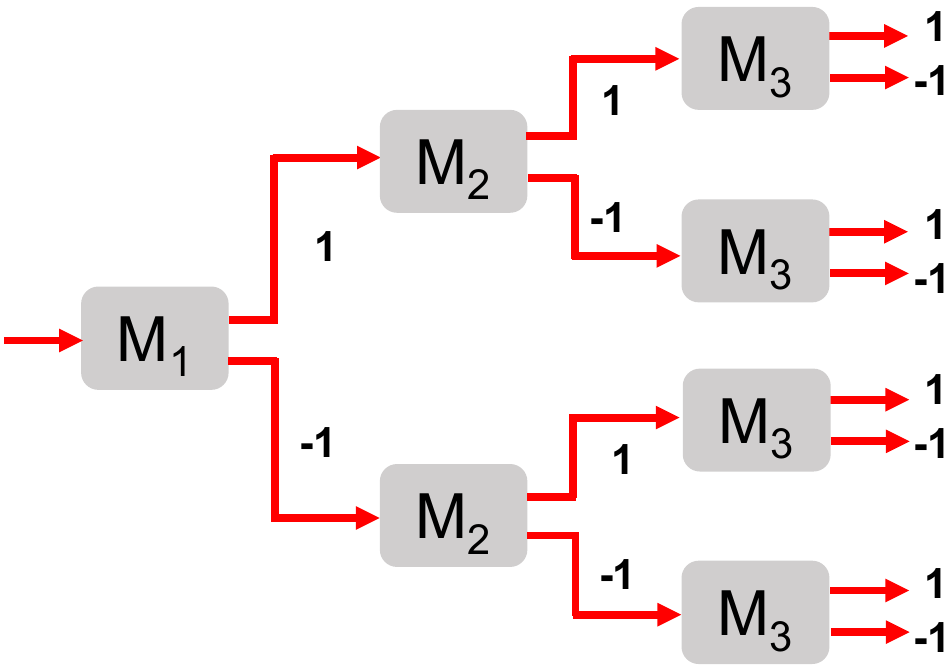}
  \caption{Devices for measuring six sets of nine observables to test inequality (\ref{eq:ine}).}
  \label{fig:2}
\end{figure}

\begin{figure*}
  \includegraphics[width=0.8\textwidth]{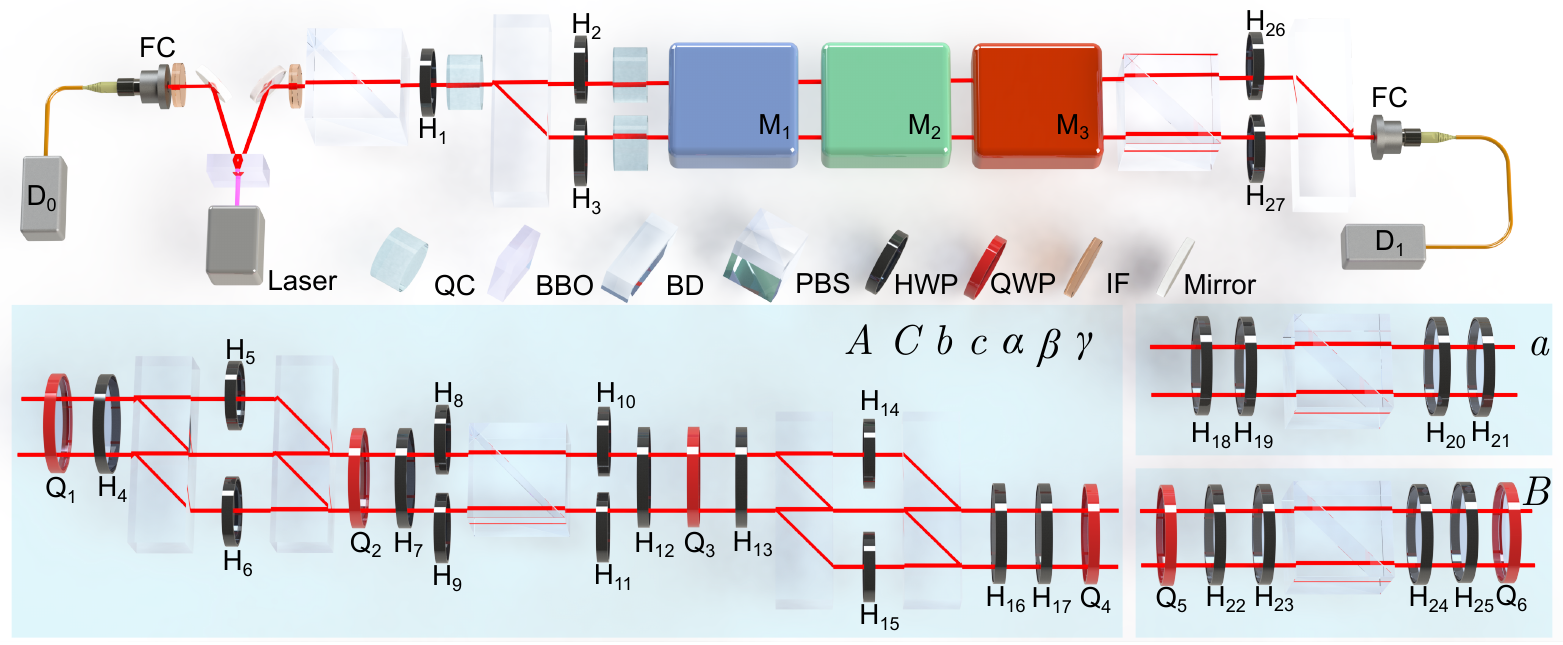}
  \caption{Experimental setup. The heralded single photons are created via type-I spontaneous parametric down-conversion
in a $\beta$-barium-borate (BBO) nonlinear crystal and are injected into the optical network (see figure for acronyms). The first polarizing beam
splitter (PBS), half-wave plates (HWPs) and beam displacer (BD) are used to generate the pure qudit states. To prepare mixed states, quartz crystals (QCs) are inserted to destroy spatial coherence of the photons~\cite{WKZ+17,WWZ+18}. The measurements are realized by wave plates and BDs. The photons are detected by APDs. The measurements $A$, $C$, $b$, $c$, $\alpha$, $\beta$ and $\gamma$ can be realized by the setup involving a PBS, four BDs and several wave plates. Whereas, the setups for realizing $a$ and $B$ can be simplified.}
  \label{fig:3}
\end{figure*}

\begin{table*}
\caption{Experimental results of each term of the entropic inequality for $26$ different states being tested. Here SD is abbreviation of standard deviation. Error bars indicate the statical uncertainty which is obtained based on assuming Poissonian statistics.}\label{tab:3}
\begin{tabular}{c||c|c|c|c|c|c|c}
  \hline\hline

   state & $H(A\cdot a\cdot\alpha)$ & $H(B\cdot b\cdot\beta)$ & $H(C\cdot c\cdot\gamma)$ & $H(A\cdot B\cdot C)$ & $H(a\cdot b\cdot c)$ & $H(\alpha\cdot\beta\cdot\gamma)$ & SD\\
  \hline\hline
  $\ket{\Psi_1}=\ket{0}$ & 0.03441(243) & 0.04503(267) & 0.05584(290) & 0.04615(271) & 0.06022(298) & 0.99988(1) & 124 \\
  \hline
  $\ket{\Psi_2}=\ket{1}$ & 0.04155(259) & 0.04921(277) & 0.04537(269) & 0.03879(253) & 0.05758(292) & 0.99993(1) & 127 \\
  \hline
  $\ket{\Psi_3}=\ket{2}$ & 0.05467(287) & 0.06471(304) & 0.04739(272) & 0.06575(307) & 0.05731(292) & 0.99991(1) & 108 \\
  \hline
  $\ket{\Psi_4}=\ket{3}$ & 0.03834(253) & 0.05590(291) & 0.04805(272) & 0.05295(285) & 0.03568(244) & 0.99990(1) & 128 \\
  \hline
  $\ket{\Psi_5}=(\ket{0}+\ket{1})/\sqrt{2}$ & 0.05146(281) & 0.06297(303) & 0.05843(295) & 0.04875(275) & 0.05298(285) & 0.99977(2) & 113 \\
  \hline
  $\ket{\Psi_6}=(\ket{0}+\ket{2})/\sqrt{2}$ & 0.03811(252) & 0.06311(305) & 0.05173(282) & 0.06159(299) & 0.05473(286) & 0.99974(3) & 115 \\
  \hline
  $\ket{\Psi_7}=(\ket{0}+\ket{3})/\sqrt{2}$ & 0.05447(286) & 0.03891(253) & 0.05249(282) & 0.06657(309) & 0.05798(292) & 0.99993(1) & 114 \\
  \hline
  $\ket{\Psi_8}=(\ket{1}+\ket{2})/\sqrt{2}$ & 0.05135(282) & 0.04230(260) & 0.05907(293) & 0.06717(307) & 0.07477(321) & 0.99992(1) & 108 \\
  \hline
  $\ket{\Psi_9}=(\ket{1}+\ket{3})/\sqrt{2}$ & 0.05254(281) & 0.05410(287) & 0.06923(310) & 0.03832(253) & 0.05248(284) & 0.99990(1) & 116 \\
  \hline
  $\ket{\Psi_{10}}=(\ket{2}+\ket{3})/\sqrt{2}$ & 0.04617(271) & 0.04837(276) & 0.05828(293) & 0.05866(294) & 0.04302(263) & 0.99984(2) & 119 \\
  \hline
  $\ket{\Psi_{11}}=(\ket{0}+\ket{1}+\ket{2})/\sqrt{3}$ & 0.07233(317) & 0.06788(311) & 0.04988(277) & 0.05100(279) & 0.07440(321) & 0.99978(2) & 102 \\
  \hline
  $\ket{\Psi_{12}}=(\ket{0}+\ket{1}+\ket{3})/\sqrt{3}$ & 0.05357(287) & 0.07477(320) & 0.06086(296) & 0.04883(276) & 0.06966(314) & 0.99992(1) & 103 \\
  \hline
  $\ket{\Psi_{13}}=(\ket{0}+\ket{2}+\ket{3})/\sqrt{3}$ & 0.05905(296) & 0.06018(297) & 0.06758(311) & 0.05434(286) & 0.07742(321) & 0.99988(1) & 101 \\
  \hline
  $\ket{\Psi_{14}}=(\ket{1}+\ket{2}+\ket{3})/\sqrt{3}$ & 0.05293(284) & 0.05973(295) & 0.06863(312) & 0.05473(287) & 0.07041(314) & 0.99992(1) & 104 \\
  \hline
  $\ket{\Psi_{15}}=(\ket{0}+\ket{1}+\ket{2}+\ket{3})/2$ & 0.05151(282) & 0.05975(298) & 0.05688(290) & 0.03355(242) & 0.03762(251) & 0.99984(2) & 124 \\
  \hline
  $\rho_{16}=(\ket{0}\bra{0}+\ket{1}\bra{1})/2$ & 0.05109(280) & 0.05873(295) & 0.06309(302) & 0.08482(331) & 0.05938(294) & 0.99983(2) & 101 \\
  \hline
  $\rho_{17}=(\ket{0}\bra{0}+\ket{2}\bra{2})/2$ & 0.05583(289) & 0.04845(274) & 0.05832(294) & 0.06057(298) & 0.06926(311) & 0.99981(2) & 108 \\
  \hline
  $\rho_{18}=(\ket{0}\bra{0}+\ket{3}\bra{3})/2$ & 0.06085(298) & 0.05591(289) & 0.05506(290) & 0.05101(281) & 0.06898(312) & 0.99982(2) & 108 \\
  \hline
  $\rho_{19}=(\ket{1}\bra{1}+\ket{2}\bra{2})/2$ & 0.06748(311) & 0.06960(312) & 0.05140(281) & 0.07073(315) & 0.07581(322) & 0.99987(2) & 96 \\
  \hline
  $\rho_{20}=(\ket{1}\bra{1}+\ket{3}\bra{3})/2$ & 0.05623(291) & 0.06646(308) & 0.07303(317) & 0.06852(310) & 0.08076(329) & 0.99978(2) & 94 \\
  \hline
  $\rho_{21}=(\ket{2}\bra{2}+\ket{3}\bra{3})/2$ & 0.06200(302) & 0.06095(297) & 0.07231(315) & 0.06083(299) & 0.05911(295) & 0.99985(2) & 101 \\
  \hline
  $\rho_{22}=(\ket{0}\bra{0}+\ket{1}\bra{1}+\ket{2}\bra{2})/3$ & 0.05917(295) & 0.06265(301) & 0.05262(283) & 0.07067(313) & 0.06372(301) & 0.99984(2) & 103 \\
  \hline
  $\rho_{23}=(\ket{0}\bra{0}+\ket{1}\bra{1}+\ket{3}\bra{3})/3$ & 0.04312(263) & 0.05509(288) & 0.06152(298) & 0.08014(329) & 0.05773(292) & 0.99979(2) & 107 \\
  \hline
  $\rho_{24}=(\ket{0}\bra{0}+\ket{2}\bra{2}+\ket{3}\bra{3})/3$ & 0.05330(285) & 0.06296(300) & 0.07202(316) & 0.06612(307) & 0.06202(301) & 0.99976(3) & 101 \\
  \hline
  $\rho_{25}=(\ket{1}\bra{1}+\ket{2}\bra{2}+\ket{3}\bra{3})/3$ & 0.05467(287) & 0.06193(301) & 0.08139(329) & 0.05878(296) & 0.06957(314) & 0.99981(2) & 99 \\
  \hline
  $\rho_{26}=(\ket{0}\bra{0}+\ket{1}\bra{1}+\ket{2}\bra{2}+\ket{3}\bra{3})/4$ & 0.07377(321) & 0.06236(301) & 0.07873(325) & 0.07640(324) & 0.07578(318) & 0.99974(3) & 89 \\
  \hline
\end{tabular}
\end{table*}

Now we present an entropic version of the state-independent contextuality proof commonly known as the Peres-Mermin square~\cite{M93,P90,M90}. It is derived for a four-level system that can be represented as a composition of two qubits that are in the same place, or even encoded on the same system. Therefore, non-locality is of no importance in this case as it is not in a non-local Bell scenario. There are nine binary $\pm1$ observables that can be measured on this system. Based on the compatibility relations, one performs these measurements in the following triples: \begin{equation}
\{A,a,\alpha\}, \{B,b,\beta\}, \{C,c,\gamma\}, \{A,B,C\}, \{a,b,c\},  \{\alpha,\beta,\gamma\}
\label{eq:measurement}
\end{equation} as shown in Fig.~\ref{fig:1}. The classical reasoning based on noncontextuality hypothesis implies that we have the distribution $\prod_{i=1}^6q_i=1$ for measured products
\begin{align} \label{qdistr}
&q_1=A\cdot a\cdot\alpha, ~~q_2=B\cdot b\cdot\beta, ~~q_3=C\cdot c\cdot \gamma,\\ \nonumber
&q_4=A\cdot B\cdot  C, ~~q_5=a\cdot b\cdot c, ~~q_6=\alpha\cdot\beta\cdot\gamma.
\end{align}
In quantum theory, we take
\begin{align}
&A=X\otimes \one, ~~a=\one\otimes X, ~~\alpha=X\otimes X, \\ \nonumber
&B=\one\otimes Y, ~~b=Y\otimes \one, ~~\beta=Y\otimes Y, \\ \nonumber
&C=X\otimes Y, ~~c=Y\otimes X, ~~\gamma=Z\otimes Z,
\end{align}
where $X$, $Y$ and $Z$ are Pauli operators. In quantum theory $\prod_{i=1}^6 q_i=-1$ as $q_1=\cdots=q_5=1$ and $q_6=-1$ for any quantum state. On the other hand, in any non-contextual realistic theory (NRT), i.e., a theory where the outcomes of $A$, $B$, etc., are predetermined (realism) and do not depend on the context in which they are measured (non-contextuality), one has $\prod_{i=1}^6 q_i=1$. This is because each such outcome appears exactly twice in two different products $q_i$ and $q_j$. Interestingly, in both NRT and quantum theory , we have $H(q_i)=0$, since the products are well defined.

The entropic non-contextual inequality derived in~\cite{RKK15} reads
\begin{align}
\label{eq:ine}
&H(\alpha\cdot\beta\cdot\gamma)\leq H(A\cdot a\cdot\alpha)+H(B\cdot b\cdot\beta)\\ \nonumber &+H(A\cdot B\cdot C)+H(a\cdot b\cdot c)+H(C\cdot c\cdot\gamma),
\end{align}
where
\begin{widetext}
\begin{align}
H(X_i\cdot X_j\cdot X_k)=-P(x_i\cdot x_j\cdot x_k=-1)\log_2 P(x_i\cdot x_j\cdot x_k=-1)
-P(x_i\cdot x_j\cdot x_k=1)\log_2 P(x_i\cdot x_j\cdot x_k=1)
\label{eq:H}
\end{align}
\end{widetext}
is the Shannon entropy of the probability distribution associated with the measurements $X_i$, $X_j$ and $X_k$ and the corresponding outcomes $x_i$, $x_j$ and $x_k$. The distributions of outcomes in both quantum and NRT do not violate the inequality. However, equal mixing of the quantum and NRT distributions maximally violates the inequality for {\it any} quantum states.

\begin{figure*}[htb]
  \includegraphics[width=0.75\textwidth]{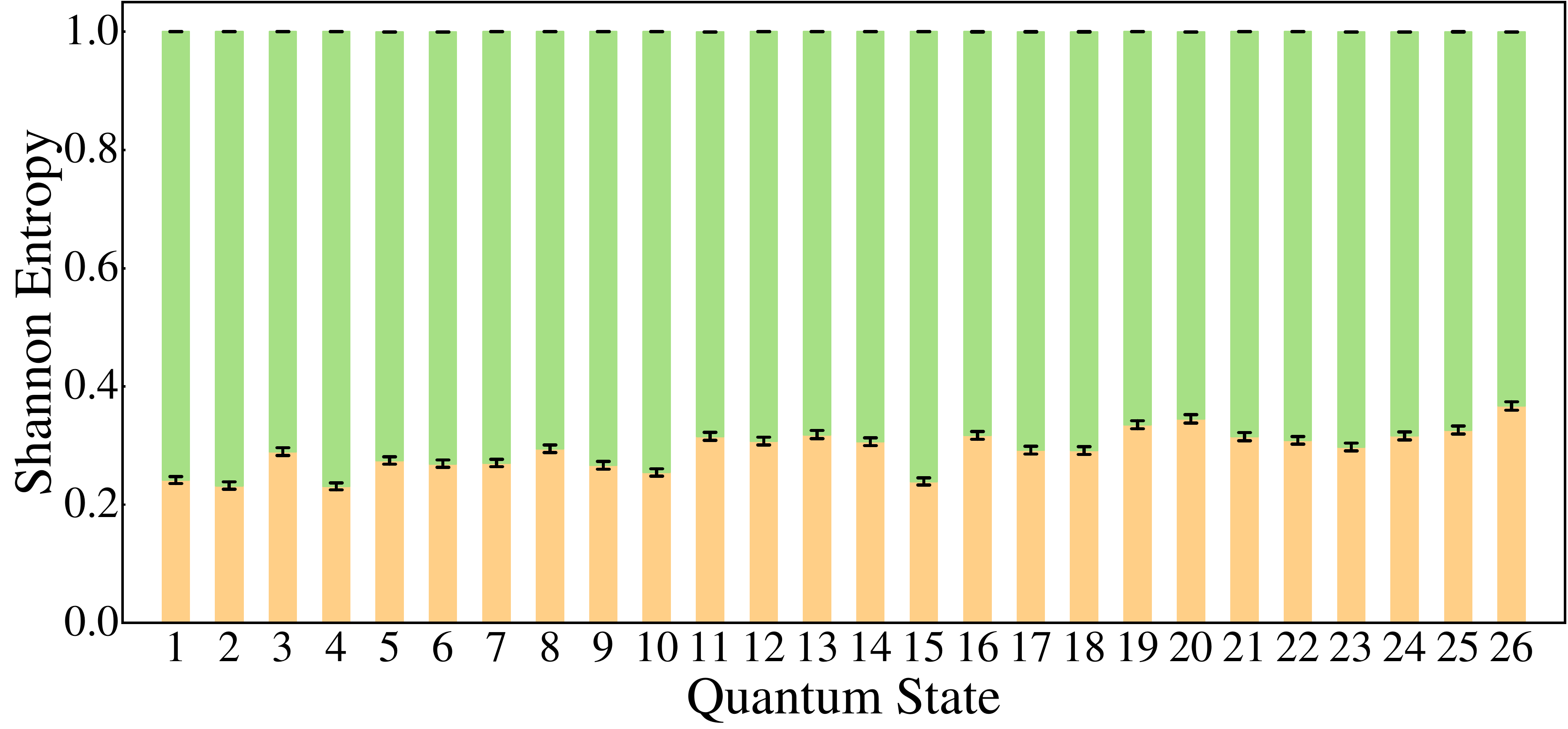}
  \caption{State-independent violation of the entropic inequality (\ref{eq:ine}). The inequality is tested for $26$ different quantum states. Left-hand side and right-hand side of the inequality are shown in green and orange bars, respectively. Error bars indicate the statical uncertainty which is obtained based on assuming Poissonian statistics.}
  \label{fig:4}
\end{figure*}


We now demonstrate experimental test of the entropic state-independent non-contextuality inequality. The purpose of the experiment is to test different quantum states of a single-particle system. In the experiment we use a single photon that simulates two qubits encoded in different degrees of freedom -- polarisation and propagation modes~\cite{WKZ+17,WWZ+18}. The basis for our two qubits is encoded as $\{\ket{0}=\ket{UH},\ket{1}=\ket{UV},\ket{2}=\ket{DH},\ket{3}=\ket{DV}\}$, where $U$ and $D$ indicate the upper and lower spatial modes of single photon, whereas $H$ and $V$ denote the horizontal and vertical polarizations of single photons. The photon pairs are generated in the spontaneous parametric down-conversion process, where one of the photons is a {\it trigger} heralding an arrival of the signal photon that we use to test the inequality. A polarization beam splitter (PBS), a beam displacer (BD) and half-wave plates (HWPs) are used to prepare the photonic four-level system in $26$ different quantum states ready for testing.

Sequential measurements of three compatible observables on the same photon are shown in Fig.~{\ref{fig:2}}. $M_i$ ($i=1,2,3$) describes the setup for measuring one of the nine observables. After the preparation stage, the photons enter the device $M_1$ through the input and yield one of the two possible outcomes.
Next, the photon enters devices $M_2$, then $M_3$ and finally it is detected at one of the eight outputs.

The experimental setup is shown in Fig.~\ref{fig:3}. Observables $a=\one\otimes X$ and $B=\one\otimes Y$ are simply rotations on the polarizations of photons keeping the spatial mode unchanged. Observables $X$ and $Y$ can be written as $M=\sum_{i=H,V} m_i\ket{m_i}\bra{m_i}$, where $\ket{m_i}$ is an eigenstate of $M$ and $m_i$ is the corresponding eigenvalue. A polarization rotation is defined $U_{M}=\ketbra{H}{m_H}+\ketbra{V}{m_V}$ and is applied on the polarization of the photons, which can be implemented by wave plates at certain setting angles following by a polarizing beam splitter (PBS). The overlap between the initial state and $\ket{m_i}$ equals to the probability of the photons being measured in the basis state $\ket{i}\in\{\ket{H},\ket{V}\}$.

Measurements $A=X\otimes\one$ and $b=Y\otimes\one$ are performed only on the spatial modes. First we use beam displacers (BDs) to split and then combine the photons with certain polarizations into the same spatial mode, which amounts to a basis transformation between spatial and polarization modes. Polarization rotations are done via wave plates following by a PBS as mentioned above. The measurements of the other observables $\alpha$, $\beta$, $C$, $c$ and $\gamma$, which are the products of two Pauli operators, are implemented by a polarization rotation, a basis transformation between spatial and polarization modes, and another polarization rotation followed by a projective measurement in the $\{\ket{H},\ket{V}\}$ basis.

We need to apply sequential measurements of three compatible observables on the same photon. Before the next measurement is done, we need to bring back the eigenstates of the previously measured observable. For six sets of measurements shown in (\ref{eq:measurement}), each set has eight different outcome distributions. For example, for the measurement $\{A,a,\alpha\}$, the eight different outcomes include $\{1,1,1\},\{1,1,-1\}$, $\{1,-1,1\}$, $\{1,-1,-1\}$, $\{-1,1,1\}$, $\{-1,1,-1\}$, $\{-1,-1,1\}$ and $\{-1,-1,-1\}$. A proper choice of devices and the wave plates' angles (please find the angles in the supplemental material), we can implement six sets of measurements with eight different outcomes. Finally, photons are detected by single-photon avalanche photodiode (APD). We only register coincidences between APD (D$_1$) and the trigger APD (D$_0$). For each outcome distribution of each measurement, we recorded clicks for $2$s, and registered about $20000$ single photons. The probability for more than one photon pair is less than $10^{-4}$ and thus it can be neglected. The coincidence counts are used to calculate the measured probabilities of eight outcomes of the each set of measurements.

In principle, for noncontextuality entropic inequalities only provide a necessary but not sufficient criterion~\cite{C13}. However, entropic inequalities turn also to be sufficient, since any contextual probabilistic model will display entropic violations if properly mixed with a classical model. To find such a non-contextual and realistic distribution, we first define
\begin{align}
\alpha=A\cdot a, \beta=B\cdot b, C=A\cdot B,
c=a\cdot b, \gamma=A\cdot B\cdot a\cdot b.
\label{eq:classical}
\end{align}
Under this definition, we have the classical distribution satisfying $q'_1=\cdots=q'_6=1$ which is analogous to (\ref{qdistr}), but takes into account the definition (\ref{eq:classical}). With the similar experimental setup, we can implement four set of measurements $\{A,a\}$, $\{B,b\}$, $\{A,B\}$ and $\{a,b\}$ and obtain the classical distribution. By equally mixing the quantum and classical distributions one obtains a distribution $\tilde{q}_i = \frac{1}{2}(q_i + q'_i)$. Note that quantum theory predicts
\begin{equation}
H(\tilde{q}_1)=\ldots = H(\tilde{q}_5)= 0,~~H(\tilde{q}_6)=1,
\end{equation}
therefore for this mixed distribution the entropic inequality is maximally violated ($1\leq 0$).

The experimental results are shown in Fig.~\ref{fig:4}. We repeat the experiment on $26$ qudit states including four basis states of four-level systems $\{\ket{\Psi_1}=\ket{0},\ket{\Psi_2}=\ket{1},\ket{\Psi_3}=\ket{2},\ket{\Psi_4}=\ket{3}\}$, six pure states with two-component superpositions of basis vectors $\{\ket{\Psi_5}=(\ket{0}+\ket{1})/\sqrt{2},\ket{\Psi_6}=(\ket{0}+\ket{2})/\sqrt{2},\ket{\Psi_7}=(\ket{0}+\ket{3})/\sqrt{2},\ket{\Psi_8}=(\ket{1}+\ket{2})/\sqrt{2},\ket{\Psi_9}=(\ket{1}+\ket{3})/\sqrt{2},\ket{\Psi_{10}}=(\ket{2}+\ket{3})/\sqrt{2}\}$, four pure states with three-component superpositions of basis vectors $\{\ket{\Psi_{11}}=(\ket{0}+\ket{1}+\ket{2})/\sqrt{3},\ket{\Psi_{12}}=(\ket{0}+\ket{1}+\ket{3})/\sqrt{3},\ket{\Psi_{13}}=(\ket{0}+\ket{2}+\ket{3})/\sqrt{3},\ket{\Psi_{14}}=(\ket{1}+\ket{2}+\ket{3})/\sqrt{3}\}$, a pure state with equal superposition of all basis vectors $\ket{\Psi_{15}}=(\ket{0}+\ket{1}+\ket{2}+\ket{3})/2$, and mixed states with different components $\rho_{16}=(\ket{0}\bra{0}+\ket{1}\bra{1})/2$, $\rho_{17}=(\ket{0}\bra{0}+\ket{2}\bra{2})/2$, $\rho_{18}=(\ket{0}\bra{0}+\ket{3}\bra{3})/2$, $\rho_{19}=(\ket{1}\bra{1}+\ket{2}\bra{2})/2$, $\rho_{20}=(\ket{1}\bra{1}+\ket{3}\bra{3})/2$, $\rho_{21}=(\ket{2}\bra{2}+\ket{3}\bra{3})/2$, $\rho_{22}=(\ket{0}\bra{0}+\ket{1}\bra{1}+\ket{2}\bra{2})/3$, $\rho_{23}=(\ket{0}\bra{0}+\ket{1}\bra{1}+\ket{3}\bra{3})/3$, $\rho_{24}=(\ket{0}\bra{0}+\ket{2}\bra{2}+\ket{3}\bra{3})/3$, $\rho_{25}=(\ket{1}\bra{1}+\ket{2}\bra{2}+\ket{3}\bra{3})/3$, and $\rho_{26}=(\ket{0}\bra{0}+\ket{1}\bra{1}+\ket{2}\bra{2}+\ket{3}\bra{3})/4$. The results in Table~\ref{tab:3} show that a state-independent violation of the entropic noncontextuality inequality (\ref{eq:ine}) occurs by $89$ standard deviations (at least). Due to the imperfections in the experiment such as the accuracy of waveplates and decoherence ect., there is a bit of difference between experimental results and theoretical predictions. On the other hand, the derivation of entropy near probability $0$ or $1$ is infinite, which
makes the entropy extremely sensitive to the noise of probability when it should be $0$ theoretically. There is also a bit of difference between the score for different states (pure states and mixed states). That is because compared to pure states, generation of desired mixed states is a little complicated and the unideal fidelities of the states influence the experimental results as well.

In our experiment photon loss opens up a detection efficiency loophole. Fair-sampling assumption is taken here which assumes that the event selected out is an unbiased representation of the whole sample~\cite{ZKK+17,LLS+11,ZZL+16}.

In summary, we experimentally demonstrate the first entropic test of state-independent contextuality on a single photonic four-level system. We show that $26$ different single photonic states violate an entropic inequality which involves correlations between results of sequential compatible measurements by at least $89$ standard deviations. Our results show that, even for a single system, and independent of its state, there is a universal set of tests whose results do not admit a noncontextual interpretation.

\begin{acknowledgements}
This work has been supported by the Natural Science Foundation of China (Grant Nos. 11674056 and U1930402) and the startup fund from Beijing Computational Science Research Centre. DK is supported by the National Research Foundation, Prime Minister's Office, Singapore and the Ministry of Education, Singapore under the Research Centres of Excellence programme. PK is supported by the National Science Centre in Poland (NCN project 2016/23/G/ST2/04273). SR is supported by the research grant system of Sharif University of Technology (G960219).
\end{acknowledgements}

\clearpage

\begin{widetext}

\appendix

\section{details of the configurations of the optical circuits for the measurements}

The details of the configurations of the optical circuits for the measurements $A$, $a$, $\alpha$, $B$, $b$, $\beta$, $C$, $c$ and $\gamma$ are shown in Tables~\ref{tab:1} and \ref{tab:2}. 

\begin{table*}[b]
\caption{Setting angles of the wave plates for realizing the measurements $A$, $\alpha$, $b$, $\beta$, $C$, $c$ and $\gamma$.}\label{tab:1}
\begin{tabular}{c||cccccccccccccccccc}
  \hline\hline
  measurement, outcome & Q$_1$ & H$_4$ & H$_5$ & H$_6$ & Q$_2$ & H$_7$ & H$_8$ & H$_9$ & H$_{10}$ & H$_{11}$ & H$_{12}$ & Q$_3$ & H$_{13}$ & H$_{14}$ & H$_{15}$ & H$_{16}$ & H$_{17}$ & Q$_4$\\
  \hline\hline
  $A,+1$ &  &  & $45^\circ$ & $45^\circ$ &  & $22.5^\circ$ & $0^\circ$ & $0^\circ$ & $0^\circ$ & $0^\circ$ & $22.5^\circ$ &  & $45^\circ$ & $45^\circ$ & $45^\circ$ & $45^\circ$ &  & \\
  $A,-1$ &  &  & $45^\circ$ & $45^\circ$ &  & $22.5^\circ$ & $45^\circ$ & $45^\circ$ & $45^\circ$ & $45^\circ$ & $22.5^\circ$ &  & $45^\circ$ & $45^\circ$ & $45^\circ$ & $45^\circ$ &  & \\
  \hline
  $\alpha,+1$ &  & $22.5^\circ$ & $45^\circ$ & $45^\circ$ &  & $22.5^\circ$ & $45^\circ$ & $0^\circ$ & $45^\circ$ & $0^\circ$ & $22.5^\circ$ &  & $45^\circ$ & $45^\circ$ & $45^\circ$ & $45^\circ$ & $22.5^\circ$ & \\
  $\alpha,-1$ &  & $22.5^\circ$ & $45^\circ$ & $45^\circ$ &  & $22.5^\circ$ & $0^\circ$ & $45^\circ$ & $0^\circ$ & $45^\circ$ & $22.5^\circ$ &  & $45^\circ$ & $45^\circ$ & $45^\circ$ & $45^\circ$ & $22.5^\circ$ & \\
  \hline
  $b,+1$ &  &  & $45^\circ$ & $45^\circ$ & $0^\circ$ & $-22.5^\circ$ & $0^\circ$ & $0^\circ$ & $0^\circ$ & $0^\circ$ & $-22.5^\circ$ & $90^\circ$ & $45^\circ$ & $45^\circ$ & $45^\circ$ & $45^\circ$ &  & \\
  $b,-1$ &  &  & $45^\circ$ & $45^\circ$ & $0^\circ$ & $-22.5^\circ$ & $45^\circ$ & $45^\circ$ & $45^\circ$ & $45^\circ$ & $-22.5^\circ$ & $90^\circ$ & $45^\circ$ & $45^\circ$ & $45^\circ$ & $45^\circ$ &  & \\
  \hline
  $\beta,+1$ & $0^\circ$ & $-22.5^\circ$ & $45^\circ$ & $45^\circ$ & $0^\circ$ & $-22.5^\circ$ & $45^\circ$ & $0^\circ$ & $45^\circ$ & $0^\circ$ & $-22.5^\circ$ & $90^\circ$ & $45^\circ$ & $45^\circ$ & $45^\circ$ & $45^\circ$ & $-22.5^\circ$ & $90^\circ$ \\
  $\beta,-1$ & $0^\circ$ & $-22.5^\circ$ & $45^\circ$ & $45^\circ$ & $0^\circ$ & $-22.5^\circ$ & $0^\circ$ & $45^\circ$ & $0^\circ$ & $45^\circ$ & $-22.5^\circ$ & $90^\circ$ & $45^\circ$ & $45^\circ$ & $45^\circ$ & $45^\circ$ & $-22.5^\circ$ & $90^\circ$ \\
  \hline
  $C,+1$ & $0^\circ$ & $-22.5^\circ$ & $45^\circ$ & $45^\circ$ &  & $22.5^\circ$ & $45^\circ$ & $0^\circ$ & $45^\circ$ & $0^\circ$ & $22.5^\circ$  & & $45^\circ$ & $45^\circ$ & $45^\circ$ & $45^\circ$ & $-22.5^\circ$ & $90^\circ$ \\
  $C,-1$ & $0^\circ$ & $-22.5^\circ$ & $45^\circ$ & $45^\circ$ &  & $22.5^\circ$ & $0^\circ$ & $45^\circ$ & $0^\circ$ & $45^\circ$ & $22.5^\circ$  & & $45^\circ$ & $45^\circ$ & $45^\circ$ & $45^\circ$ & $-22.5^\circ$ & $90^\circ$ \\
  \hline
  $c,+1$ &  & $22.5^\circ$ & $45^\circ$ & $45^\circ$ & $0^\circ$ & $-22.5^\circ$ & $45^\circ$ & $0^\circ$ & $45^\circ$ & $0^\circ$ & $-22.5^\circ$ & $90^\circ$ & $45^\circ$ & $45^\circ$ & $45^\circ$ & $45^\circ$ & $22.5^\circ$ &  \\
  $c,-1$ &  & $22.5^\circ$ & $45^\circ$ & $45^\circ$ & $0^\circ$ & $-22.5^\circ$ & $0^\circ$ & $45^\circ$ & $0^\circ$ & $45^\circ$ & $-22.5^\circ$ & $90^\circ$ & $45^\circ$ & $45^\circ$ & $45^\circ$ & $45^\circ$ & $22.5^\circ$ &  \\
  \hline
  $\gamma,+1$ &  & $0^\circ$ & $45^\circ$ & $45^\circ$ &  & $0^\circ$ & $45^\circ$ & $0^\circ$ & $45^\circ$ & $0^\circ$ & $0^\circ$ &  & $45^\circ$ & $45^\circ$ & $45^\circ$ & $45^\circ$ & $0^\circ$ &  \\
  $\gamma,-1$ &  & $0^\circ$ & $45^\circ$ & $45^\circ$ &  & $0^\circ$ & $0^\circ$ & $45^\circ$ & $0^\circ$ & $45^\circ$ & $0^\circ$ &  & $45^\circ$ & $45^\circ$ & $45^\circ$ & $45^\circ$ & $0^\circ$ &  \\
  \hline
\end{tabular}
\end{table*}

\begin{table*}[b]
\caption{Setting angles of the wave plates for realizing the measurements $a$ and $B$.}\label{tab:2}
\begin{tabular}{c||cccccccccc}
  \hline\hline
  measurement, outcome & H$_{18}$ & H$_{19}$ & H$_{20}$ & H$_{21}$ & Q$_5$ & H$_{22}$ & H$_{23}$ & H$_{24}$ & H$_{25}$ & Q$_6$ \\
  \hline\hline
   $a,+1$ & $22.5^\circ$ & $0^\circ$ & $0^\circ$ & $22.5^\circ$ &  &  &  &  &  &  \\
   $a,-1$ & $22.5^\circ$ & $45^\circ$ & $45^\circ$ & $22.5^\circ$ &  &  &  &  &  &  \\
   \hline
   $B,+1$ &  &  &  &  & $0^\circ$ & $-22.5^\circ$ & $0^\circ$ & $0^\circ$ & $-22.5^\circ$ & $90^\circ$ \\
   $B,-1$ &  &  &  &  & $0^\circ$ & $-22.5^\circ$ & $45^\circ$ & $45^\circ$ & $-22.5^\circ$ & $90^\circ$ \\
   \hline
\end{tabular}
\end{table*}

\end{widetext}


\begin{references}
\bibitem{S60} E. P. Specker, Dialectical anthropology {\bf14}, 239 (1960).
\bibitem{B66} J. S. Bell, Rev. Mod. Phys. {\bf38}, 447 (1966).
\bibitem{KS67} S. Kochen and E. P. Specker, J. Math. Mech. {\bf17}, 59 (1967).
\bibitem{M93} N. D. Mermin, Rev. Mod. Phys. {\bf65}, 803 (1993).
\bibitem{C08} A. Cabello, Phys. Rev. Lett. {\bf101}, 210401 (2008).
\bibitem{BBCP09} P. Badzi\c{a}g, I. Bengtsson, A. Cabello and I. Pitowsky, Phys. Rev. Lett. {\bf103}, 050401 (2009).
\bibitem{BB84} C. H. Bennett and G. Brassard. ``Quantum cryptography: Public key distribution and coin tossing''. In Proceedings of IEEE International Conference on Computers, Systems and Signal Processing, volume 175, page 8. New York (1984).
\bibitem{Shor} P. W. Shor, SIAM J. Sci. Statist. Comput. {\bf26}, 1484 (1997).
\bibitem{NC00} M. A. Nielsen and I. L. Chuang, ``Quantum Computation and Quantum Information,'' Cambridge University Press, Cambridge (2000).
\bibitem{C13} R. Chaves, Phys. Rev. A {\bf87}, 022102 (2013).

\bibitem{KRK12} P. Kurzy\'{n}ski, R. Ramanathan and D. Kaszlikowski, Phys. Rev. Lett. {\bf109}, 020404 (2012).
\bibitem{CF12} R. Chaves and T. Fritz, Phys. Rev. A {\bf85}, 032113 (2012).
\bibitem{CF13} T. Fritz and R. Chaves, IEEE Trans. Inf. Theory {\bf59}, 803 (2013).
\bibitem{ZKK+17} X. Zhan, P. Kurzy\'{n}ski, D. Kaszlikowski, K. K. Wang, Z. H. Bian, Y. S. Zhang and P. Xue, Phys. Rev. Lett. {\bf 119}, 220403 (2017).
\bibitem{RKK15} S. Raeisi, P. Kurzy\'{n}ski and D. Kaszlikowski, Phys. Rev. Lett. {\bf 114}, 200401 (2015).
\bibitem{ARBC09} E. Amselem, M. R{\aa}dmark, M. Bourennane and A. Cabello, Phys. Rev. Lett. {\bf 103}, 160405 (2009).
\bibitem{KZG+09} G. Kirchmair F. Z\"{a}hringer, R. Gerritsma, M. Kleinmann, O. G\"{u}hne, A. Cabello, R. Blatt and C. F. Roos, Nature {\bf 460}, 494 (2009).
\bibitem{ZWD+12} C. Zu, Y.-X. Wang, D.-L. Deng, X.-Y. Chang, K. Liu, P.-Y. Hou, H.-X. Yang and L.-M. Duan, Phys. Rev. Lett. {\bf 109}, 150401 (2012).
\bibitem{P90} A. Peres, Phys. Lett. A {\bf151}, 107 (1990).
\bibitem{M90} N. D. Mermin, Phys. Rev. Lett. {\bf65}, 3373 (1990).
\bibitem{WKZ+17} K. K. Wang, G. C. Knee, X. Zhan, Z. H. Bian, J. Li and P. Xue, Phys. Rev. A {\bf 95}, 032122 (2017).
\bibitem{WWZ+18} K. K. Wang, X. P. Wang, X. Zhan, Z. H. Bian, J. Li, B. C. Sanders and P. Xue, Phys. Rev. A {\bf 97}, 042112 (2018).
\bibitem{LLS+11} R. Lapkiewicz, P. Li, C. Schaeff, N. K. Langford, S. Ramelow, M. Wiesniak and A. Zeilinger, Nature (London) {\bf474}, 490 (2011).
\bibitem{ZZL+16} X. Zhan, X. Zhang, J. Li, Y. S. Zhang, B. C. Sanders and P. Xue, Phys. Rev. Lett. {\bf116}, 090401 (2016).

\end{references}
\end{document}